\documentclass[10pt,a4paper,twocolumn]{article}
\usepackage{eurosis}
\usepackage{graphicx}
\usepackage{url}
\usepackage{times}
\usepackage{authblk}

\title{Performance analysis of a 240 thread tournament level MCTS Go program on
  the Intel Xeon Phi}
\author[1,2]{Ali Mirsoleimani\thanks{email:
    s.a.mirsoleimani@liacs.leidenuniv.nl}}
\author[1]{Aske Plaat}
\author[2]{Jos Vermaseren}
\author[1]{Jaap  van den Herik}
\affil[1]{Leiden Centre of Data Science, Leiden University, The
  Netherlands}
\affil[2]{Nikhef Theory Group, Nikhef Amsterdam,  The Netherlands}

\date{}

\begin{document}
\maketitle
\thispagestyle{empty}


\begin{abstract}
In 2013 Intel introduced the Xeon Phi, a new parallel co-processor
board. The Xeon Phi is a
cache-coherent many-core shared memory architecture claiming
CPU-like versatility, programmability, high performance, and power
efficiency. The first published
micro-benchmark studies indicate that many of Intel's claims appear
to be true. The current paper is the first study on the Phi of a
complex artificial intelligence application. It contains an open source
MCTS application for playing tournament quality Go (an
oriental board game). We report the first
speedup figures for up to 240 parallel threads on a real
machine, allowing a direct comparison to previous simulation
studies. After a substantial amount of work, we observed that performance scales
well up to 32 threads, largely confirming previous simulation 
results of this Go program, although the performance surprisingly deteriorates between
32 and 240 threads. Furthermore, we report (1) unexpected performance anomalies
between the Xeon Phi and Xeon CPU for small problem sizes and small
numbers of threads, and (2) that performance is 
sensitive to scheduling choices. Achieving good performance on the
Xeon Phi for complex programs is not straightforward; it requires a
deep understanding of (1) search patterns, (2) of scheduling,
and (3) of the architecture and its many cores and caches. In
practice, the
Xeon Phi is less straightforward to program for than originally envisioned
by Intel. 
\end{abstract}

\section{Introduction}
In 2013 Intel introduced a new many-core architecture, a cache-coherent shared memory co-processor architecture claiming
CPU-like versatility, programmability, high performance, and power
efficiency (in contrast to hard-to-program energy-hungry GPU
co-processor architectures from companies such as NVIDIA). When a new
architecture emerges there is a great interest in the community for
analyzing and understanding its performance. When GPGPU
  programming became prevalent, a rich
literature on GPU performance modeling and simulations emerged. See, e.g.,
~\cite{Karami, mirsoleimani2014two}, and~\cite{Lopez-Novoa2014a}. This
trend is now starting for the Xeon Phi. The first published
micro-benchmark studies indicate that many of Intel's claims are true
\cite{Fang2014}. Of course, 
micro-benchmarks only tell a part of the story and performance of
actual applications may differ in practice. In this paper we
have chosen to study an important application from the domain of
artificial intelligence.  

Ever since the victory of IBM's {\sc Deep Blue} over World Chess Champion
Garry Kasparov on 11 May 1997 \cite{SchaefferPlaatICCA:1997}, computer
Go has been the {\em Drosophila Melanogaster\/} of Artificial
Intelligence. The complexity and depth of the game has
frustrated AI researchers trying to replicate the computer chess
successes for many years \cite{Herik2013}. The brute-force minimax
approach, so successful in chess, turned out to be a dead end in Go,
when in 2006 a new probabilistic simulation algorithm was
introduced. This new algorithm, Monte Carlo Tree Search
\cite{Chaslot2008,Coulom2006,Kocsis2006,Gelly2011} was successful,
beating the first human Go-grandmaster in 2008. Not only was MCTS 
successful in Go, it also proved successful in many other
combinatorial optimization and simulation
problems~\cite{Kuipers2013,Ruijl2014,Browne2012}.  

Because of the successes with MCTS, much research effort has been put
into improving the performance of parallel MCTS
algorithms. MCTS performance studies have become important in their own
right in recent years
\cite{Chaslot2008,Yoshizoe2011a,Bourki2011,Segal:2010:SPU:1950322.1950326,Schaefers2014}. However,
scaling studies of MCTS on shared memory many-core machines have been
limited to smaller studies, typically 8-24 core machines
\cite{Chaslot2008}, or have been performed in simulation
\cite{Segal:2010:SPU:1950322.1950326}. The advent of the Intel Xeon
Phi allows for the first time to replicate such simulation studies on
actual hardware, up to 240 simultaneous threads. We have performed
this study using {\sc Fuego} \cite{Enzenberger2010a}, the same open source
program that was used in the simulation study \cite{Segal:2010:SPU:1950322.1950326},
allowing  a direct comparison. 

This paper has two main contributions.
\begin{itemize}
\item We have performed,  to the best of our
knowledge, the first performance study of a non-trivial MCTS
program on the Intel Xeon Phi.  We find unexepected sensitivity to
problem size and scheduling, which we attribute to a low integer performance
and complex interconnect architecture.
\item We have performed the first large scale (up to 240 threads) study of MCTS
  tree parallelism on a real shared memory many core machine. We find good
  performance up to 32 threads, confirming a previous simulation
  study, and deteriorating performance from 32 to 240
  threads.
\end{itemize}

Moreover, we have two more findings. First, Intel's wish of high performance at no cost to the
programmer is only partly achieved, due to the complex hardware
characteristics of the Xeon Phi architecture. Second, the scaling of the
algorithm is dependent on many details including cache hierarchy,
access latency, scheduling policy, and core architecture. 

The remainder of this paper is structured as follows: in section 2 the
architecture of the Xeon Phi is briefly discussed. Section 3 discusses
related work. Section 4 gives the experimental setup, and section 5
gives the experimental results. Finally, scalability and thread
affinity are discussed, and a conclusion is given. 




\section{Architecture of Intel Xeon Phi Co-processor}
\begin{figure}
\centering
\includegraphics[scale=0.35]{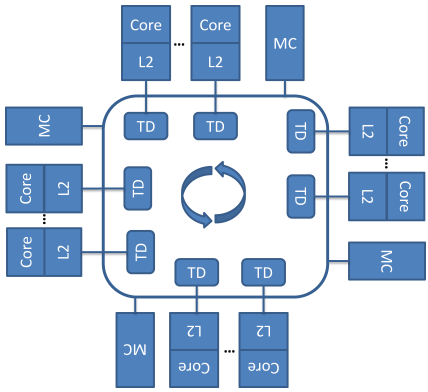}
\caption{Intel Xeon Phi microarchitecture}
\label{fig:phi}
\end{figure}

We start providing an overview  of the Intel Xeon Phi co-processor
architecture (see Figure~\ref{fig:phi}). A Xeon Phi co-processor board
consists of up to 61 cores based on the Intel 64-bit ISA. Each
  of these cores contains \textit{vector processing units} (VPU) to
  execute 512 bits of 8 double-precision floating point elements or 16
  single-precision floats or 32-bit integers at the same time, 4-way
  SMT, and dedicated L1 and fully coherent L2
  caches~\cite{Rahman2013}.  The \textit{tag directories} (TD) are
  used to look up cache data distributed among the cores. The
  connection between cores and other functional units such as
  \textit{memory controllers} (MC) is through a bidirectional
  \textit{ring interconnect}. There are 8 distributed memory
  controllers as interface between the ring burst and main memory
  which is up to 16 GB.  

\begin{figure}
\centering
\includegraphics[scale=0.6]{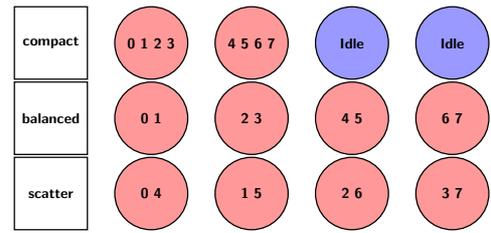}
  \caption{Allocation of 8 threads on a 4 core system with 4 threads per core with different affinity types.}
  \label{fig:ER-10}
  \end{figure}

The scheduling policy on the Xeon Phi can be
influenced manually at run time. By setting an environment variable
one can control how the threads are bound to cores. This can be
advantageous for exploiting data locality of the algorithm. Using the
\textit{KMP\_AFFINITY} environmental variable threads can be
distributed among cores. Possible settings are:
\textit{compact}, \textit{balanced}, or \textit{scatter}. The
\textit{compact} type allocates threads to cores in a way that
maximizes cache utilization while \textit{scatter} type do thread
allocation to maximize core utilization. Figure \ref{fig:ER-10} shows
the allocations of threads for different types. If threads access data
that is stored in a cache nearby, the \textit{balanced} type is the best choice
because it maximizes cache and core utilization simultaneously.

\section{Related Work}
\begin{figure}
\centering
\includegraphics[scale=0.8]{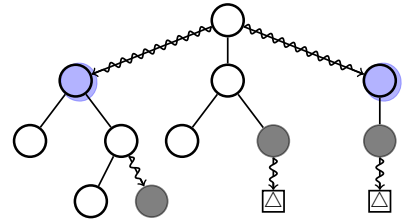}
\caption{Tree parallelization with local lock. The curly arrows represent threads. The shadowy nodes are locked ones. The black nodes are newly added to the tree.}
\label{fig:PMCTS-1}
\end{figure}

Below we review related work on MCTS parallelizations.
The four major parallelization methods for MCTS are leaf
parallelization, root parallelization, tree
parallelization~\cite{Chaslot2008}, and transposition table driven
work scheduling (TDS) based approaches~\cite{Yoshizoe2011a}. 
Of these, tree parallelization is the method most often used on shared
memory machines. It is the method used in {\sc Fuego}. In tree
parallelization one MCTS tree is shared among several threads that are 
performing simultaneous searches~\cite{Chaslot2008}. The main
challenge in this method is using data locks to prevent data
corruption. Figure \ref{fig:PMCTS-1} shows the tree parallelization
algorithm with local locks. A lock-free implementation of this
algorithm addressed the aforementioned problem with better
scaling than a locked approach~\cite{Enzenberger2010a}.  
There is also a case study that shows a good   performance of a
(non-MCTS) Monte Carlo simulation on the Xeon Phi
co-processor~\cite{Shuo-li2013}.  


Schaefers et al.~\cite{Schaefers2014} propose a parallel MCTS method
for distributed memory systems called \textit{UCT-Treesplit}. Yoshizoe
et al.~\cite{Yoshizoe2011a} describe a parallelization approach based
on TDS~\cite{Romein99,Romein2002} for MCTS called
\textit{depth-first UCT}. There are some attempts to parallelize MCTS
on accelerator processors including GPU~\cite{Rocki2011}. 

Segal reports the scaling of tree parallelization with virtual loss in
{\sc Fuego} for different number of threads and time controls on a
simulated idealized shared-memory
system~\cite{Segal:2010:SPU:1950322.1950326} . He finds that strength
of play increases asymptotically with as resources increase (more time
or more threads). A near-perfect speedup is reported for 64 threads
and 60-minute per game. Segal suggests that speedup starts decreasing
beyond 64 threads, although, with large time settings, further scaling
to 512 threads still shows performance increases.

Enzenberger et al.\cite{Enzenberger2010b} evaluate tree parallelization with virtual loss and
local locks on a 16-core shared-memory system. The algorithm shows an
eight-fold speedup with 16 threads.

\section{Experimental Setup: Xeon CPU against Xeon Phi}
To determine the effective performance on the Xeon Phi we have performed self-play
experiments. A major problem with parallel game playing programs is
the phenomenon of {\em search overhead}, which occurs since, in
parallel, parts of the search tree will be searched that a
sequential search would already have found to be unimportant. Therefore a simple
{\em efficiency\/} measure such as games per second will not tell the whole
story. In order to determine the {\em effective speedup\/} a different
method is needed. We have performed self-play experiments in which a
version of the program with double the resources (2x \# of threads) against a
version with single resources (1x \# of threads) are compared \cite{Chaslot2008}. 

In order to generate statistically significant results in a reasonable
amount of time most works use the setting of 1 second per move, and so
did we, initially. 
{\sc Fuego} is an open source, tournament level Go playing program,
developed by a team headed by  Martin Mueller at the University
of Alberta~\cite{Enzenberger2010a}. The experiments were conducted with
{\sc Fuego} SVN revision 1900, on a 9x9 board, with komi 6, 
Chinese rules, alternating player color was enabled, opening book was
disabled. The win-rate of two opponents is measured by running at least a 100-game match. A single
game of Go typically lasts 200 moves. The games were played using the
Gomill python library for tournament play~\cite{Woodcraft2014}.

A statistical method based on ~\cite{Heinz2001} is used to calculate
95\%-level confidence lower and upper bounds on the real winning rate
of a player. Assume $p$ is the true winning probability of a
player. The value of $p$ is estimated by $0 \leq w=x/n \leq 1$ which
results from $x\leq n$ wins in a match of $n$ games. Therefore, we may
simply assume $w$ the sample mean of a binary-valued random variable
that counts two draws as a loss plus a win. 
The expected value of $w$ is $E(w)=p$ and the variance of $w$ is
$Var(w)=p(1-p)/n$. According to the central limit theorem
approximately, $w \approx Normal(p, p(1-p)n)$, so
$(w-p)/\sqrt{p(1-p)/n}\approx Normal(0,1)$. Let $z_{\%}$ denote the
upper critical value of the standard $N(0,1)$ normal distribution for
any desired \%-level of statistical confidence($z_{90\%}=1.645$,
$z_{95\%}=1.96$). Then, the probability of $w-1.96\sqrt{p(1-p)/n}\leq
p \leq w+1.96\sqrt{p(1-p)/n}$ is about 95\%. Therefore, the 95\%
confidence interval on the true winning probability $p$ is
$[w-1.96\sqrt{p(1-p)/n}, w+1.96\sqrt{p(1-p)/n}]$. The value of unknown
$p$ is substituted for $w$: $[w-1.96\sqrt{w(1-w)/n},
w+1.96\sqrt{w(1-w)/n}]$. 

Our Xeon Phi co-processor board is hosted on a machine in which a standard
12-core Xeon CPU is present. This allows experiments in which a
conventional parallel Xeon CPU architecture is pitted against the new parallel Xeon
Phi architecture. 

The results were measured on a machine with (1) an Intel {\em Xeon
  CPU\/} E5-2695 2.40GHz with 12 cores and 48 hyperthreads. Each
physical core has 256KB L2 cache and the chip has a total of 30MB L3
cache. The machine has 160GB physical memory. (2) An Intel {\em Xeon
  Phi\/} 7120P 1.238GHz is installed which has 61 cores and 244
hardware threads. Each core has 512KB L2 cache. The co-processor has
16GB GDDR5 memory on board with an aggregate theoretical bandwidth of
352 GB/s. The peak  turbo frequency is 1.33GHz. The theoretical performance of the 7120P is
2.416 TFLOPS or TIPS and 1.208 TFLOPS for single-precision or integer and
double-precision floating-point arithmetic operations,
respectively~\cite{Intel2013}.

Intel's {\em icc 14.0.1} compiler is used to compile {\sc Fuego} in {\em native mode}. A {\em native application} runs directly on the Xeon Phi and its embedded Linux operating system.
\section{Experimental Results}
\begin{figure}[t]
\centering
\includegraphics[scale=0.5]{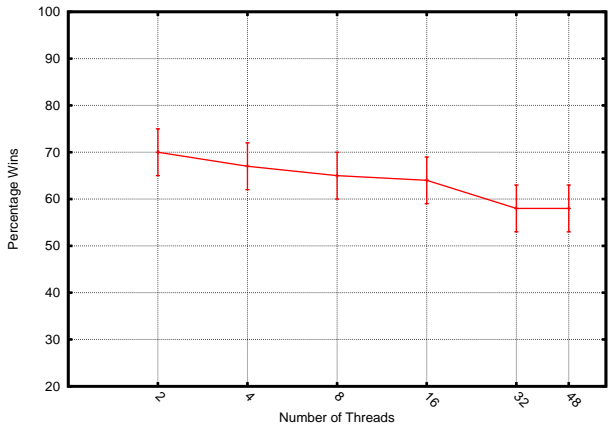}
\caption{Performance of self-play {\sc Fuego} with n threads against {\sc Fuego}
  with n/2 threads on the Xeon CPU processor. 200 games
  for each data point. }
\label{fig:ER-4}
\end{figure}

{\em Xeon CPU\/}: Figure \ref{fig:ER-4} shows the results of the self-play experiments
for {\sc Fuego} on the conventional {\em Xeon CPU}.  For the 9x9 board the win-rate of
the program with double the number of threads is better than the base
program, starting at 70\%, decreasing to 58\% at 32 threads and then
becomes flat. (The 19x19 board has a similar performance, not shown). These results are entirely in
line with results reported in ~\cite{Enzenberger2010b} for 16 vs 8 threads. The
slightly decreasing lines are explained by the phenomenon of search overhead: the
parallel program with double the amount of threads (e.g., 16 threads) searches more parts
of the tree than the version of the program with half the number
of threads (e.g., 8 threads).

\begin{figure}[t]
\centering
\includegraphics[scale=0.5]{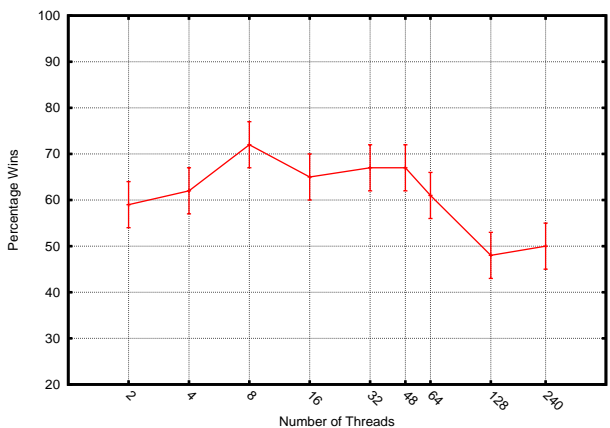}
\caption{Performance of self-play {\sc Fuego} on the Xeon Phi with n threads against {\sc Fuego}
  with n/2 threads. The board size is 9x9. 300 games for each data point.}
\label{fig:ER-2}
\end{figure}

{\em Xeon Phi\/}: Figure \ref{fig:ER-2} shows our initial results for the win-rate on the {\em
  Xeon Phi}. For these experimental settings (1 second per move) the Phi-graph differs markedly
from the CPU-graph in Figure \ref{fig:ER-4}. The Xeon CPU shows a smooth,
slightly decreasing line. The Xeon Phi shows a more ragged line that first
slopes up, and then slopes down. Also, the overall win-rate is
lower than on the CPU, and even dips below 50\%, implying that more threads
actually loses from less threads! (all within margins of error as
indicated by the error bars). The best win-rate on the Xeon Phi is
for 8 threads while on the Xeon CPU it is on 2 threads. The playing strength remains
above the break-even point of 50\% for the first player until 48
threads and then sharply decreases, until 
128 threads and becomes 50 percent for 240 threads. 
Up to 64 threads, these results basically confirm the simulation study
by Segal~\cite{Segal:2010:SPU:1950322.1950326}. However, beyond 64
threads the performance drop is unexpectedly large.

The most striking feature of the experiment with these settings is the difference in 
performance of an identical program in an identical setup on the Xeon CPU 
versus the Xeon Phi: Figures \ref{fig:ER-4} and \ref{fig:ER-2} do not
look alike at all. The Xeon CPU shows a steadily decreasing performance, as 
expected, where the Xeon Phi shows a ragged hump-like shape.

To study the possible causes of these results, we performed 
experiments to delve deeper into the Xeon Phi architecture.


\subsection{Efficiency, Thread Affinity, and Problem Size}
\begin{figure}[t]
\centering
\includegraphics[scale=0.5]{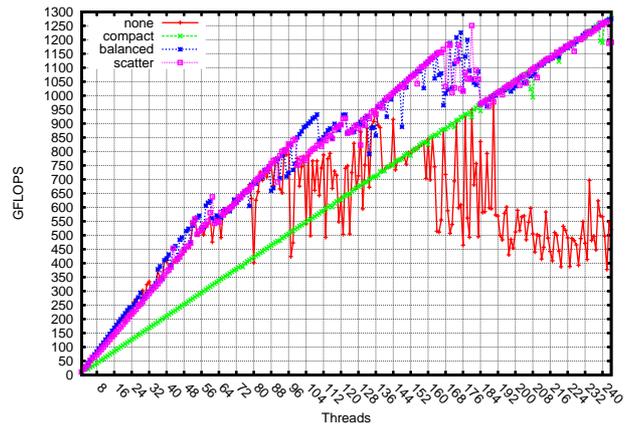}
\caption{Performance of double-precision operations of the Xeon Phi for different numbers of threads.}
\label{fig:ER-9}
\end{figure}

\begin{figure}[t]
\centering
\includegraphics[scale=0.5]{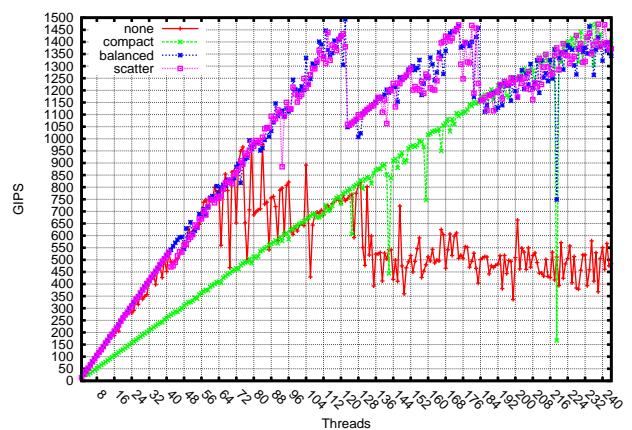}
\caption{Performance of integer operations of the Xeon Phi for
  different numbers of threads.}
\label{fig:ER-12}
\end{figure}

\begin{figure}[t]
\centering
\includegraphics[scale=0.5]{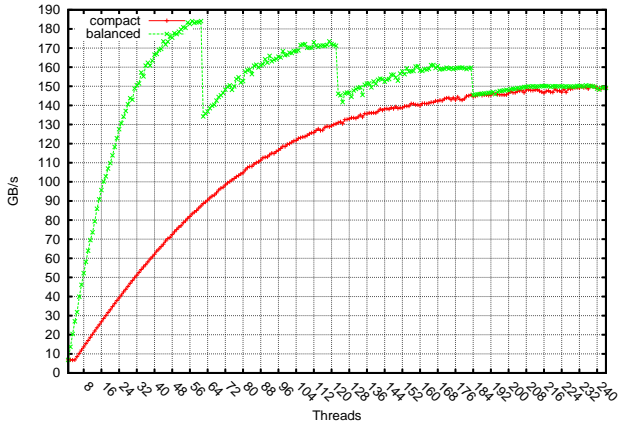}
\caption{Memory bandwidth of the Xeon Phi for different numbers of threads.}
\label{fig:ER-11}
\end{figure}

As mentioned before, the thread scheduling policy on the Xeon Phi can
be influenced manually at run time. In order to illustrate graphically
the effect of different scheduling policies we have performed a small
experiment. Figure \ref{fig:ER-9} and \ref{fig:ER-12} show the effect
of different thread affinities on the performance of the Xeon Phi for
double-precision and integer arithmetic operations. The benchmark is
executing a loop which contains $c[j]=a[j]*b[j]+c[j]$  operation for
many times. The  effect of thread affinities on the bandwidth of the
Xeon phi for executing the same program in double-precision is also
shown in Figure \ref{fig:ER-11}. The results were measured with turbo
mode set to on. 

In the \textit{compact} mode the performance was steadily increased and the bandwidth reached a plateau. In the \textit{balanced} and \textit{scatter} modes depending on how many threads are assigned to each core 4 different
regions for double and 3 different region for integer performance existed. For example, as shown in Figure \ref{fig:ER-12} between 122 threads and 183 threads some cores have 2 threads and some others have three threads in \textit{balanced} mode. This asymmetry degraded the performance dramatically at the beginning of the region and then stared to increase performance. The memory bandwidth has also 4 regions in \textit{balanced} mode. By using more thread the bandwidth never reached the same level as the previous region. These type of performance behavior makes it really tricky to select best thread configuration for executing a program like {\sc Fuego}.

\begin{figure}
\centering
\includegraphics[scale=0.5]{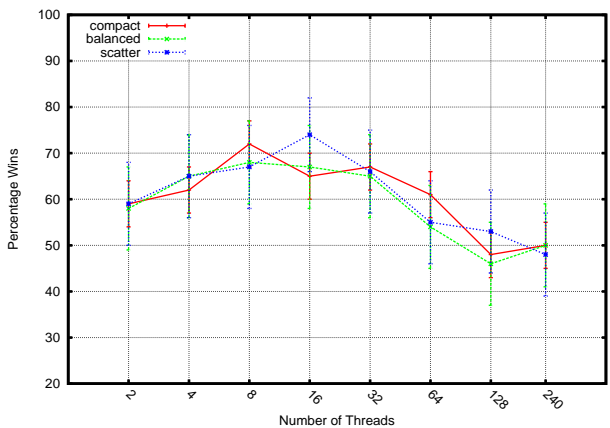}
\caption{Self-play performance of {\sc Fuego} for different thread
  affinity types on the Xeon Phi. The board size is 9x9. 100 games for each
  data point. } 
\label{fig:ER-8}
\end{figure}

For the {\sc Fuego} self-play experiments the {\em compact} affinity type has been used. To show the effect of different scheduling policies on {\sc Fuego} the three different methods have been run. Figure \ref{fig:ER-8} shows the effect of different thread affinity types on the performance of {\sc Fuego}. The percentage of wins for \textit{balanced} mode shows more stability compared to the two other scheduling methods. The best win-rate is for 4 threads (1 core) in \textit{compact} mode and for 16 threads (16 cores) for \textit{scatter} mode. 

As noted before, the most striking feature of these experiments is the
difference in performance of an identical program in an identical
setup on the Xeon CPU versus the Xeon Phi using the standard
experimental settings of the 9x9 board and 1 second per move. The Xeon CPU shows a steadily decreasing performance, as expected, where the Xeon Phi shows a ragged hump-like shape. 

\begin{figure}[t]
\centering
\includegraphics[scale=0.5]{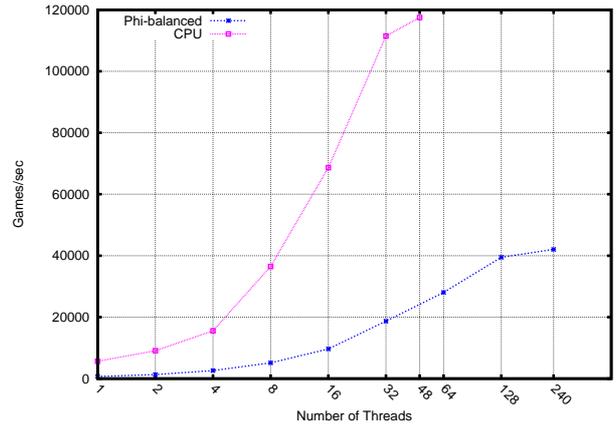}
\caption{Number of games per second for a 9x9 board when {\sc Fuego} makes the second move.}
\label{fig:ER-7}
\end{figure}

\begin{figure}[t]
\centering
\includegraphics[scale=0.5]{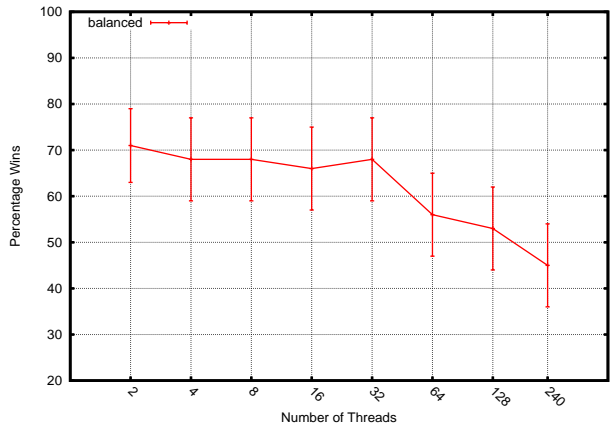}
\caption{Self-play performance of {\sc Fuego} for 10 second per move on the Xeon Phi. The board size is 9x9. 100 games for each
  data point.}
\label{fig:ER-13}
\end{figure}

The win-rate graph shows an {\em effective speedup}, a speedup measure
that includes search overhead. We have also computed basic {\em efficiency}-speedup figures to compare the parallel speed of the Xeon CPU and the Xeon Phi. The {\sc Fuego} program can output the number of games per second that are performed. Figure
\ref{fig:ER-7} shows the number of games per second that is performed
by {\sc Fuego} on a 9x9 board. This is a convenient measure of how
efficient the architecture is running the program. Games per second is
the number of games that are played by {\sc Fuego} before making a
move. The number increases for both architectures. Due to the higher
clock speed, the amount of work by the each core of the Xeon CPU is
much more than the Xeon Phi core. However, the difference in clock
speed is only a factor of two, whereas in the figure the difference
is more than a factor of 5. Figure \ref{fig:ER-7} shows that even
using all cores of the Xeon Phi cannot reach the performance of 16
threads on The Xeon CPU. The low games-per-second numbers of the 
Xeon Phi suggests inefficiencies due to a small problem size. Closer
inspection of the results on which Figure
\ref{fig:ER-7} is based suggests that {\sc Fuego} is not able to do enough
simulation on the Xeon Phi for small number of threads in just 1
second. Therefore, we increase the time limit per move to 10
seconds. Figure \ref{fig:ER-13} shows the results of the self-play 
experiment when the {\sc Fuego} can make a move with 10 seconds for
doing simulation on The Xeon Phi. We see that now the graph is approaching
that of the Xeon CPU. The win-rate behavior for low number
of threads is now much closer to that of the CPU
(Figure~\ref{fig:ER-2}), and the counter-intuitive hump-shape has
changed to the familiar down-sloping trend. However, we still see fluctuation in  the
\textit{balanced} mode. Up to 32 threads performance is still reasonable (close
to 70\% win-rate for the 2x thread program) but up to 240 threads performance deteriorates. The maximum win-rate is for 8 threads and there is still a marginal benefit for using 128 threads. 
 
\begin{figure}[t]
\centering
\includegraphics[scale=0.5]{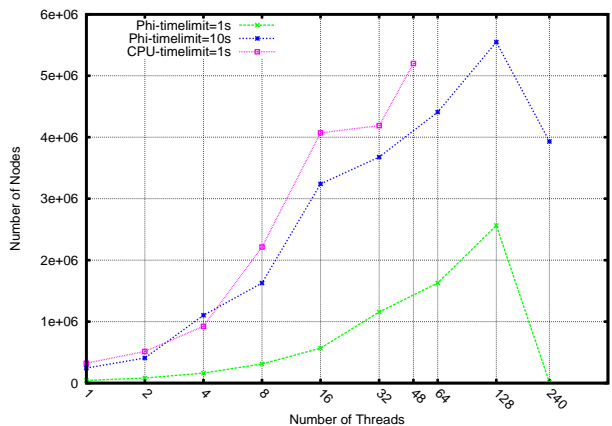}
\caption{The number of nodes in the search tree when {\sc Fuego} makes the second move. \textit{balanced} thread
  affinity type on the Xeon Phi is used. The board size is 9x9.}
\label{fig:ER-14}
\end{figure}

The reason behind the difference between the results in
  Figures \ref{fig:ER-2} and \ref{fig:ER-13} is shown in Figure
  \ref{fig:ER-14} which shows how large the search tree is when making a move. The size of the tree when {\sc Fuego} has 10 seconds per move on the Xeon Phi is similar to  Xeon CPU with 1 second per move.
\section{Conclusion}
Intel's Xeon Phi has been designed to offer high performance without the
associated hassle of difficult programming. To this end, the Xeon Phi
architecture has many cores, many levels of caches, an intricate
cache-coherency protocol, vector units, and
a fast and complex interconnect. A complex machinery to offer a simple
programming model. Previously,
micro-benchmarks have been reported that indicate that Intel's
published figures are essentially achieved in practice~\cite{Fang2014}. 

Using a tournament quality game playing program that employs a popular Monte Carlo
method~\cite{Browne2012}, our results show that porting a
complex program and getting it to run correctly on the Xeon Phi is indeed relatively
straightforward. Our results are, to our knowledge, the first
performance results of a non-trivial AI program on an shared
memory architecture with up to 240 threads, allowing comparison of a
simulation study up to 64 threads~\cite{Segal:2010:SPU:1950322.1950326} and an extension
beyond. 

However, the results also show that achieving good performance is not straightforward: on the
conventional Xeon CPU architecture our results are as predicted, a
smooth, slightly down-sloping line, while, using the standard experimental settings
of 1 second per move, on the Xeon Phi we initially see a ragged hump-like
shape that differs markedly from the Xeon CPU. Our further experiments
with scheduling policies and problem sizes were prompted
by the assumption that cache locality issues and memory access patterns
caused the unexpected performance results. We find that performance is quite
sensitive to different settings, especially problem size. By judicious choice of scheduling
strategy and problem size, we were able to achieve reasonable
results after considerable analysis and tuning. Results
appear first to be largely in line with a previous simulation study, showing
reasonable scaling up to 32 threads, but deteriorating performance
up to 240 threads. 

From our experimental results we may conclude that achieving good performance on the
Xeon Phi for complex programs is not straightforward; achieving good
performance requires a deep understanding of the search algorithm on the one
hand and the architecture and its many cores and caches on the other
hand. We suggest that in this case a complex architecture does not
automatically equal a simple performance model. Given the industry
trend towards heterogeneous many-core architectures, 
increasing our understanding of the interplay between algorithm and
architecture is vital for achieving good 
performance. More research is under way
to create more accurate performance models of the Intel Xeon Phi for
different types of algorithms and MCTS-like access patterns.

\section{Acknowledgment}
This work is supported in part by the ERC Advanced Grant no. 320651, ``HEPGAME.''

\bibliographystyle{abbrv}
\bibliography{library}

\begin{thebibliography}{10}

\bibitem{Bourki2011}
A.~Bourki, G.~Chaslot, M.~Coulm, V.~Danjean, H.~Doghmen, J.-B. Hoock,
  H.~Thomas, A.~Rimmel, F.~Teytaud, O.~Teytaud, P.~Vayssi, T.~H\'{e}rault,
  P.~Vayssi\`{e}re, and Z.~Yu.
\newblock {Scalability and Parallelization of Monte-Carlo Tree Search}.
\newblock In {\em Proceedings of the 7th International Conference on Computers
  and Games}, CG'10, pages 48--58, Berlin, Heidelberg, 2011. Springer-Verlag.

\bibitem{Browne2012}
C.~B. Browne, E.~Powley, D.~Whitehouse, S.~M. Lucas, P.~I. Cowling,
  P.~Rohlfshagen, S.~Tavener, D.~Perez, S.~Samothrakis, and S.~Colton.
\newblock {A Survey of Monte Carlo Tree Search Methods}.
\newblock {\em Computational Intelligence and AI in Games, IEEE Transactions
  on}, 4(1):1--43, 2012.

\bibitem{Chaslot2008}
G.~M. J.~B. Chaslot, M.~H.~M. Winands, and H.~J. van~den Herik.
\newblock {Parallel monte-carlo tree search}.
\newblock {\em Computers and Games}, 5131:60--71, 2008.

\bibitem{Coulom2006}
R.~Coulom.
\newblock {Efficient Selectivity and Backup Operators in Monte-Carlo Tree
  Search}.
\newblock In {\em Proceedings of the 5th International Conference on Computers
  and Games}, CG'06, pages 72--83, Berlin, Heidelberg, May 2006.
  Springer-Verlag.

\bibitem{Enzenberger2010a}
M.~Enzenberger and M.~M\"{u}ller.
\newblock {A lock-free multithreaded Monte-Carlo tree search algorithm}.
\newblock {\em Advances in Computer Games}, 6048:14--20, 2010.

\bibitem{Enzenberger2010b}
M.~Enzenberger, M.~Muller, B.~Arneson, and R.~Segal.
\newblock {Fuego-An Open-Source Framework for Board Games and Go Engine Based
  on Monte Carlo Tree Search}.
\newblock {\em IEEE Transactions on Computational Intelligence and AI in
  Games}, 2(4):259--270, Dec. 2010.

\bibitem{Fang2014}
J.~Fang, A.~A.~L. Varbanescu, H.~Sips, L.~Zhang, C.~Xu, and Y.~Che.
\newblock {Test-driving Intel Xeon Phi}.
\newblock In {\em Proceedings of the 5th ACM/SPEC international conference on
  Performance engineering - ICPE '14}, number Section III, pages 137--148, New
  York, New York, USA, Mar. 2014. ACM Press.

\bibitem{Gelly2011}
S.~Gelly and D.~Silver.
\newblock {Monte-Carlo tree search and rapid action value estimation in
  computer Go}.
\newblock {\em Artificial Intelligence}, 175(11):1856--1875, 2011.

\bibitem{Heinz2001}
E.~Heinz.
\newblock {New self-play results in computer chess}.
\newblock In {\em Computers and Games}, pages 262--276, 2001.

\bibitem{Intel2013}
Intel.
\newblock {Intel® Xeon Phi Product Family Highly parallel processing to power
  your breakthrough innovations}.
\newblock
  http://www.intel.com/content/www/us/en/bench-marks/server/xeon-phi/xeon-phi-theoretical-maximums.html,
  2013.

\bibitem{Karami}
A.~Karami, S.~A. Mirsoleimani, and F.~Khunjush.
\newblock {A Statistical Performance Prediction Model for OpenCL Kernels on
  NVIDIA GPUs}.
\newblock In {\em The 17th CSI International Symposium on Computer Architecture
  \& Digital Systems (CADS'13)}, Tehran, Iran, 2013. IEEE.

\bibitem{Kocsis2006}
L.~Kocsis and C.~Szepesv\'{a}ri.
\newblock {Bandit based monte-carlo planning}.
\newblock {\em Machine Learning: ECML 2006}, 2006.

\bibitem{Kuipers2013}
J.~Kuipers, A.~Plaat, J.~Vermaseren, and H.~van~den Herik.
\newblock {Improving multivariate Horner schemes with Monte Carlo tree search}.
\newblock {\em Computer Physics Communications}, 184(11):2391--2395, Nov. 2013.

\bibitem{Lopez-Novoa2014a}
U.~Lopez-Novoa, A.~Mendiburu, and J.~Miguel-Alonso.
\newblock {A Survey of Performance Modeling and Simulation Techniques for
  Accelerator-based Computing}.
\newblock {\em IEEE Transactions on Parallel and Distributed Systems},
  9219(c):1--1, 2014.

\bibitem{mirsoleimani2014two}
S.~A. Mirsoleimani, A.~Karami, and F.~Khunjush.
\newblock {A Two-Tier Design Space Exploration Algorithm to Construct a GPU
  Performance Predictor}.
\newblock In {\em Architecture of Computing Systems--ARCS 2014}, pages
  135--146. Springer, 2014.

\bibitem{Rahman2013}
R.~Rahman.
\newblock {\em {Intel Xeon Phi Coprocessor Architecture and Tools: The Guide
  for Application Developers}}.
\newblock Apress, Sept. 2013.

\bibitem{Rocki2011}
K.~Rocki and R.~Suda.
\newblock {Large-Scale Parallel Monte Carlo Tree Search on GPU}.
\newblock In {\em Parallel and Distributed Processing Workshops and Phd Forum
  (IPDPSW), 2011 IEEE International Symposium on}, pages 2034--2037, May 2011.

\bibitem{Romein2002}
J.~Romein, H.~Bal, J.~Schaeffer, and A.~Plaat.
\newblock {A performance analysis of transposition-table-driven work scheduling
  in distributed search}.
\newblock {\em IEEE Transactions on Parallel and Distributed Systems},
  13(5):447--459, May 2002.

\bibitem{Romein99}
J.~Romein, A.~Plaat, H.~E. Bal, and J.~Schaeffer.
\newblock {Transposition Table Driven Work Scheduling in Distributed Search}.
\newblock In {\em In 16th National Conference on Artificial Intelligence
  (AAAI'99)}, pages 725--731, 1999.

\bibitem{Ruijl2014}
B.~Ruijl, J.~Vermaseren, A.~Plaat, J.~Herik, and H.~J. van~den Herik.
\newblock {Combining Simulated Annealing and Monte Carlo Tree Search for
  Expression Simplification}.
\newblock {\em Proceedings of ICAART Conference 2014}, 1(1):724--731, 2014.

\bibitem{Schaefers2014}
L.~Schaefers, M.~Platzner, and S.~Member.
\newblock {Distributed Monte-Carlo Tree Search : A Novel Technique and its
  Application to Computer Go}.
\newblock {\em IEEE Transactions on Computational Intelligence and AI in
  Games}, 6(3):1--15, 2014.

\bibitem{SchaefferPlaatICCA:1997}
J.~Schaeffer and A.~Plaat.
\newblock {Kasparov versus deep blue: The re-match}.
\newblock {\em ICCA Journal}, 20(2):95--102, 1997.

\bibitem{Segal:2010:SPU:1950322.1950326}
R.~B. Segal.
\newblock {On the Scalability of Parallel UCT}.
\newblock In {\em Proceedings of the 7th International Conference on Computers
  and Games}, CG'10, pages 36--47, Berlin, Heidelberg, 2011. Springer-Verlag.

\bibitem{Shuo-li2013}
Shuo-li.
\newblock {Case Study: Achieving High Performance on Monte Carlo European
  Option Using Stepwise Optimization Framework}.
\newblock
  https://software.intel.com/en-us/articles/case-study-achieving-high-performance-on-monte-carlo-european-option-using-stepwise,
  2013.

\bibitem{Herik2013}
H.~J. van~den Herik, A.~Plaat, J.~Kuipers, and J.~A.~M. Vermaseren.
\newblock {Connecting Sciences}.
\newblock {\em In 5th International Conference on Agents and Artificial
  Intelligence (ICAART 2013)}, 1:IS--7 -- IS--16, 2013.

\bibitem{Woodcraft2014}
M.~Woodcraft.
\newblock {Gomill library}.
\newblock http://mjw.woodcraft.me.uk/gomill/, 2014.

\bibitem{Yoshizoe2011a}
K.~Yoshizoe, A.~Kishimoto, T.~Kaneko, H.~Yoshimoto, and Y.~Ishikawa.
\newblock {Scalable Distributed Monte-Carlo Tree Search}.
\newblock {\em Fourth Annual Symposium on Combinatorial Search}, pages
  180--187, May 2011.

\end{thebibliography}
\end{document}